\documentclass[aps,showpacs,pre,onecolumn]{revtex4}
\usepackage{epsfig}
\usepackage{amsmath}
\usepackage{graphicx,natbib}%amssymb

\begin{document}
%\doublespacing
%\bibliographystyle{apsrev}%apsrev}%prsty}

\title{Ising-like dynamics in large-scale functional brain networks}

\author{Daniel Fraiman$^1$, Pablo Balenzuela$^2$, Jennifer Foss$^3$, Dante R. Chialvo$^3$}

\affiliation{$^1$ Departamento de Matem\'atica y Ciencias,
Universidad de San Andr\'es and CONICET, Buenos Aires, Argentina.}

\affiliation{$^2$ Departmento de F\'isica, Facultad de Ciencias
Exactas y Naturales, Universidad de Buenos Aires and CONICET,
Buenos Aires, Argentina.}

\affiliation{$^3$ Department of Physiology,  Northwestern
University,  Chicago, Illinois, USA. }

\begin{abstract}

Brain ``rest'' is defined -more or less unsuccessfully- as the
state in which there is no explicit brain input or output. This
work focuss on the question of whether such state can be
comparable to any known \emph{dynamical} state. For that purpose,
correlation networks from human brain Functional Magnetic
Resonance Imaging (fMRI) are constrasted with correlation networks
extracted from numerical simulations of the Ising model in 2D, at
different temperatures. For the critical temperature $T_c$,
striking similarities appear in the most relevant statistical
properties, making the two networks indistinguishable from each
other. These results are interpreted here as lending support to
the conjecture that the dynamics of the functioning brain is near
a critical point.

\end{abstract}
 \date{\today}
\maketitle

\section{INTRODUCTION}
The human cerebral cortex is organized as a very complex network
comprising approximately $10^{10}$ interconnected neurons. Thanks
to the impressive progress in brain imaging techniques given by
the development of positron emission tomography  and functional
Magnetic Resonances Imaging (fMRI) an increasing amount of
spatiotemporal brain data is now available. The analysis of this
large and complex information is of such unprecedented magnitude
that conceptual approaches grounded in statistical physics are
needed\cite{werner}.

Recently, and departing from the tradition of using
stimulus-response techniques, the study of brain imaging dynamics
``at rest'', have received ample
attention\cite{fox_07,raichle_01,raichle_06a,greicius_03}.  Brain
``rest'' is defined -more or less unsuccessfully- as the state in
which there is no explicit brain input or output. The analysis of
experiments under such quasi-stationary state revealed an active
network of brain areas engaged during the resting state in a
peculiar way. Typically, although these areas are active during
rest, they will shutdown immediately when the subject engages in
any minimal cognitive task; for example when asked to visually
track a moving object on a screen. This evidence, now expanded by
other studies, indicates the existence of a so-called brain
``resting state network'' or also ``default mode network'' in
which several cortical regions are activated on a complex
cooperative/competitive dynamical interaction.

Results from brain imaging experiments as well as graph theory
analysis already agree on some fundamental common features, which
can be summarized as follow:
\begin{enumerate}
\item{There are dense local correlations with only few long range
links, resembling a Small World network \cite{Stam,eguiluz,
vanden,Salvador,Bullmore}.}
 \item{The distribution of
the number of links is scale-free\cite{eguiluz,vanden}, when
measured with the appropriate resolution.}

\item{Brain networks are assortative, indicating a tendency for
nodes with similar number of links to be directly connected
\cite{eguiluz,SpornsPlos,Park2008}.}

\item{Large positively correlated domains coexist with equally
large anti-correlated non-local structures \cite{fox_05}.}
 \item{Large-scale correlated
patterns (or their graphs counterpart) have been observed during
subject execution of a task, as well as under ``brain rest''
conditions (absence of an overt stimulus)~\cite{fox_05,
baliki2,eguiluz} and even under general anesthesia
~\cite{vincent_07}.}

\item{A portion of these observations cannot be explained by the
brain's underlying ``anatomical'' connectivity \cite{SpornsPlos}.}
\end{enumerate}

Although there is at least one colloquial explanation for each of
the points listed above, a single mechanistic explanation that
satisfies all these observations at once is still lacking. This is
already in itself an important theoretical deficit, but it is
additionally highlighted by the fact that in certain brain
disfunctions some of these global properties are known to be
affected\cite{baliki2, Stamreview1,Stamreview2}.

%A fundamental question, yet never explored, then is to identify
%which of these spatiotemporal dynamics are trivially related with
%the brain structure (i.e., connections) and which are due to
%collective emergent dynamics.

As a starting point we ask here whether the brain resting state
could be comparable to any known \emph{dynamical} state. We have
proposed that the brain stays near the critical point of a second
order phase transition, where neuronal groups generate a diversity
of flexible collective behaviors, due to the known abundance of
metastable states at the transition. It is from this viewpoint,
that the dynamics of brain resting might correspond to a critical
state. This conjecture is tested here comparing fMRI brain resting
state data from healthy subjects with a paradigmatic critical
system, the Ising model \cite{Ising}. This model has been the
``fruitfly'' for the development of concepts and techniques in
statistical thermodynamics. It is chosen based on the qualitative
similarities between some of its dynamics and the brain's fMRI
spatiotemporal patterns which contain long-range
correlations\cite{eguiluz,baliki2,chialvo2004,chialvo2007,chialvo2008}
and a mixture of ordered and disordered structures.  It must be
noted from the outset that we are not suggesting that the brain's
equation are isomorphic with those of the Ising model.
Nevertheless, the results in this paper do suggest that important
lessons can be learned from the striking similarities between the
brain data and the dynamics emerging from the Ising model at
critical temperature. The important point is that the numerical
experiments here are not simulations prepared to replicate and
further study a given experimental finding. To the contrary, the
phenomenology to be discussed, is not written in any way in the
model equations, nevertheless all the features listed above for
the brain appear spontaneously in the Ising model near the
critical temperature.

The paper is organized as follows. The next section is dedicated
to describing the data from the brain and from the numerical
simulations of the Ising model. Also it contains the steps used to
extract the networks from both, brain and model time series.
Section III contains the main finding organized as a side-by-side
comparison of the statistical properties of the brain and Ising
model networks. It is shown that key statistical and topological
properties of the brain networks are intriguingly similar to those
of the networks extracted from the Ising (only) at the critical
temperature. Finally Section IV summarizes and discuss the
relevance of these similarities and their biological significance
in terms of brain functioning.

\section{EXPERIMENTAL AND NUMERICAL METHODS}
In this work, two types of complex networks are analyzed in
detail. The first network is derived from time series of brain
fMRI images collected from healthy human volunteers. The second
network is extracted from numerical simulations of the Ising
model\cite{Ising} in 2D. In both systems networks are defined in
the same manner by linking sites with strongest correlations,
often called ``correlation networks'', as it will be explained in
detail below.

\subsection{The brain fMRI data}
Functional magnetic resonance data was acquired using a 3T Siemens
Trio whole-body scanner with echo-planar imaging (EPI) capability
using the standard radio-frequency head coil (scanning parameters
were as in \cite{baliki2}). Data used here correspond to five
healthy females with ages ranging between 28 and 48 years old.
They were all right-handed, and all gave informed consent to
procedures approved by Northwestern University IRB committee.
Participants were scanned following a typical brain resting state
protocol\cite{fox_07}, in which the subject is lying in the
scanner and asked to keep their mind blank, eyes closed and
avoid falling asleep. A total of 300 images are obtained spaced
by 2.5 sec. in which the brain oxygen level dependent (BOLD)
signal is recorded for each one of the 64x64x49 sites (so-called
voxels of dimension 3.4375mm x 3.4375mm x 3mm). Typically, only
10\% of those voxels correspond to brain activity, the type of time
series used here. Preprocessing of BOLD signal was performed using
FMRIB Expert Analysis Tool (FEAT, \cite{jezzard},
http://www.fmrib.ox.ac.uk/fsl), involving motion correction using
MCFLIRT; slice-timing correction using Fourier-space time-series
phase-shifting; non-brain removal using BET; spatial smoothing
using a Gaussian kernel of full-width-half-maximum 5mm.

\subsection{The Ising model}

The Ising model considers a lattice containing $N$ sites and
assumes that each lattice site $i$ has an associated variable
$s_i$, where $s_i=+1$ stands for an ``up''  spin and $s_i=-1$ for
a ``down''  spin.  A particular configuration of the lattice is
specified by the set of variables ${s_1,s_2,...,s_N}$ for all
lattice sites. The energy in absence of external magnetic field is
given by
\begin{equation}
E = -J\sum_{i,j=nn(i)}^N s_is_j
\label{eq:ising}
\end{equation}
where $N=L \times L$, $L$ is the size of the lattice, $J$ is the
coupling constant, and the sum over $j$ run over the nearest
neighbors of a given site $i$ ($nn(i)$). As in almost all
statistical mechanics models there exists a competition between
thermal fluctuations (given by the interaction with the
environment) that give the system a tendency to be disordered, and
the interaction between particles (sites of the lattice) that
tends to  organize the system in some particular way that depends
on the interaction or coupling between particles.

We implement the Metropolis Monte Carlo algorithm
\cite{Metro,Tobo} for the evolution of the Ising model in 2D with
periodic boundary conditions. This algorithm takes into account
that the system is in contact with a heat bath at temperature $T$.
In this work, instead of working with asymptotic configurations
and equilibrium averages, we deal with temporal series of single
spin dynamics observed at a certain timescale. All simulations are
implemented on a lattice of $L=200$ and every time step
corresponds to $\Delta t=L \times L$ Monte Carlo steps (which
corresponds, on average, to running once over the entire lattice,
giving each spin the possibility to flip).  We take the Boltzman
constant equal $1$, $J=1$, and after thermalization, we take
$2000$ lattice configurations (each separated by $\Delta t$),
obtaining the dynamics of each spin
($\{s_1(t),s_2(t),\dots,s_N(t)\}$ time series).

As mentioned in the introduction some of the key properties
exhibited by the brain resemble the dynamics of the Ising model at
the critical temperature $T_c$, where a transition between the
ordered and disordered states takes place. At lower temperatures
almost all the spins are aligned, while at temperatures above
critical, spins are randomly distributed and the total
magnetization is approximately zero. At the critical temperature
$T_c$ however, the system displays a fractal structure, with
clusters of aligned spins of different sizes as well as long range
temporal correlations. For the simulations discussed in the
Results section a critical temperature $T_{c}=2.3$ was used,  a
subcritical one of $T=2.0$ and a supercritical temperature of
$T=3$. A final note concerns the choice of a lattice with nearest
neighbors ferromagnetic interactions, considering that the brain
is not a lattice and that includes inhibitory (i.e.,
anti-ferromagnetic) interactions. This is chosen deliberately as
the worst case scenario ir order to demonstrate the higher
significance of the critical dynamics over the structural
connectivity of the model. We restricted ourselves to the
discussion of the present Ising configuration, since the main
results are connected with the critical state itself, however they
are expected to be observed after changes in the connectivity and
type of interactions, provided that the system is tuned near the
critical state.

\subsection{Correlation Networks}

In general terms, networks are collections of nodes joined by
links. For certain systems, the nodes as well as the links are
self evident and easily identifiable. This is not the case at
hand, because in the brain both are part of the problem, where
both the \emph{nodes and their interactions} need to be uncovered
from the data. Although there are ``anatomical'' templates that
can be used to identify nodes, we choose here to approach the
problem from the limit of maximum ignorance and use instead a
data-driven strategy. The assumption is that the brain time series
contain enough information to define the networks in a
self-consistent manner. Since the correlations are computed from
the time series collected from fMRI voxels, this approach is often
called voxel-based brain functional networks. The current approach
is in contrast with most recent related
work\cite{Salvador,Bullmore,Achard,Basset}, where the network's
nodes are predefined based on a priori knowledge, and only the
possible links between these predefined nodes are determined. It
will be seen that these differences per se can be responsible for
conflicting results.

Here networks are defined by the correlations among the activity
at each location (i.e., either a voxel in the case of the brain,
or a lattice site in the Ising model). Thus the correlation
coefficient, $r$, is used to measure  the degree of linear
dependence between all pairs of sites, as was done
in~\cite{eguiluz}. The correlation coefficient between sites $i$
and $j$ is:
\begin{equation}
r(i,j)= \frac{\langle x_i(t)x_j(t)\rangle -\langle x_i(t)\rangle
\langle x_j(t)\rangle}{\sigma(x_i(t))\sigma(x_j(t))},
\label{eq:correl}
\end{equation}
where $\sigma^2(x_l(t))=\langle x_l^2(t) \rangle -\langle x_l(t)
\rangle^2$, and $x_l(t)$ is the BOLD signal of voxel $l$ if we are
studying the brain data, or the spin time series (of site $l$)
from the Ising model. $\langle  .  \rangle$ represents averages
taken over the length of the time series (300  and 2000 points for
the brain and the Ising model, respectively).

\emph{Links} between sites $i$ and $j$ are defined here whenever
the correlation $r(i,j)$ is greater or equal to a given
threshold, $\rho$. The network's \emph{nodes} are those sites with
a non-zero number of links. That completes the definition of a
network.  Depending on the sign of $\rho$, two types of networks
can be extracted. Those using a positive threshold $\rho^{+}$ will
be called positively correlated networks and those using a
negative threshold $\rho^{-}$ negatively correlated networks. It
is relevant to make this differentiation because  anti-correlated
dynamics are ubiquitous in the brain, as will be discussed latter.

The next section contains a side-by-side analysis of the
positively and negatively correlated networks extracted from the
brain and the Ising model.
\begin{figure}
\begin{center}
\includegraphics[width=0.3\textwidth,angle=0,keepaspectratio,clip]{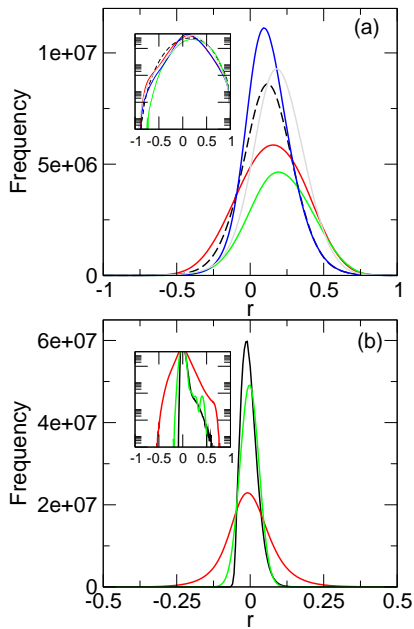}
\end{center}
\caption{Panel (a): Brain's fMRI correlation density distribution
for each of the five control subjects recorded under resting
conditions. Panel (b) Ising model correlation density distribution
for $T=2$ (green),  $T=2.3$ (red), and $T=3$ (black). Insets in
both panels show the log-linear plot of the same density
distributions.}
 \label{fig1}
 \end{figure}

\section{RESULTS}
Networks are extracted from the data using the site-to-site
temporal correlations. Fig.~\ref{fig1} shows the  density
distribution for the $N(N-1)/2$ correlations (i.e., Eq. 2)
computed between all pairs of time series of the brain (in Panel
a) and the Ising model (in Panel b). In the case of the brain, the
results from the five participants are plotted and for the Ising
model, the densities correspond to correlations computed at three
different temperatures: subcritical, critical and supercritical.

It can be seen that besides the differences in variance, which can
not be expected to be equal, the densities for both brain and
Ising model are distributed approximately equally. Both have a small
skewness towards positive values, which it is more clearly seen in
the insets using logarithmic scale. An important point to
appreciate in Panel B of Fig. 1 is the well known increase in the
variance of the correlations at the critical temperature. It is
only near $T_c$ that equally oriented spins coalesce in large
domains thus generating the two sides of the distribution we
observe here. In the brain, at any moment in time, in order to
produce a given motor or cognitive behavior, of even during rest,
similar dynamics occur: large regions of the brain activate in
bulk at the same time that other regions de-activate. A remark
here is that it is inconceivable to think about the brain's
ongoing dynamics in any other way. Given the brain extensive
connectivity, this balance in which a region is shutting down
while another is excited is clearly the only possibility to avoid
both total quiescence, in which the brain is shutdown, and
massive excitation in which the entire cortex is fired up.
Thus, while the reason for the distribution shown in
Fig. 1a is trivial, it is not trivial how the brain does it, or in other
words, which is the mechanism in place to maintain such balanced
correlations \cite{problem}.

\subsection{Networks average statistical properties}

\begin{figure}
\begin{center}
\includegraphics[angle=0,width=0.80\textwidth,keepaspectratio,clip]{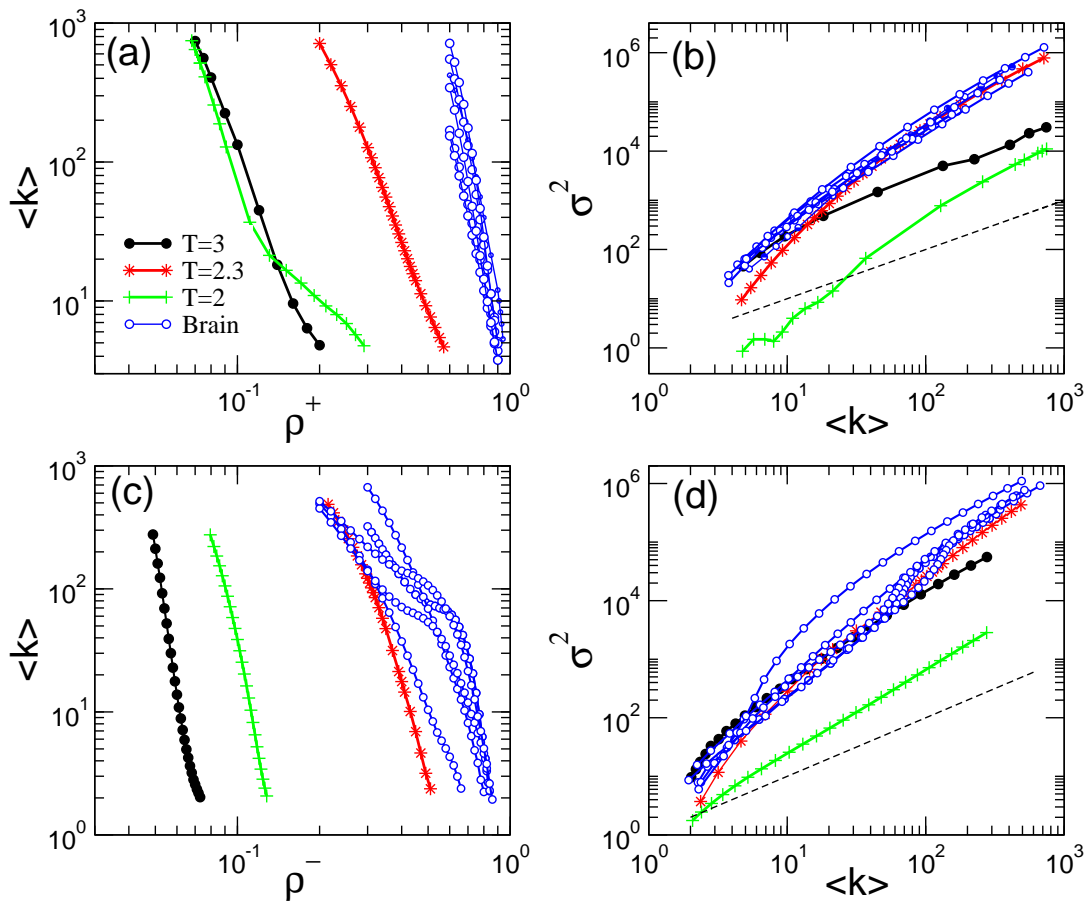}
\end{center}
\caption{Average degree, $\langle k \rangle$, as a function of
threshold $\rho$ for positive (a) and negative (c) correlation
networks. Variance of degree, $\sigma^2_k$, as a function of
$\langle K \rangle$ for positively  (b) and negatively (d)
correlated networks. The dashed black line (in panels b and d),
corresponds to the expected behavior of a Poisson distribution. In
all graphs the Ising model data at three temperatures $T=2$,
$T=2.3$, and $T=3$  is presented, as well as the data from the
five subject's fMRI brain networks.} \label{fig2}
\end{figure}

Given the differences in correlation variance discussed above, a
criterion needs to be established to compare brain and Ising
networks. One proven to be useful is to scan a range of $\rho$
thresholds while computing the average degree of the resulting
networks. After that, a comparison can be made between networks of
similar average degree. This is plotted in Fig.~\ref{fig2} as a
function of $\rho$ for positively correlated networks in panel (a)
and for negatively correlated networks in panel (c). This is done
for the brain data of the five subjects, and for the Ising data at
three temperatures. Given the distributions in Fig. 1, it is clear
that, as $\rho$ grows, there are fewer connections, and
consequently a smaller average degree, $\langle k \rangle$.
Despite the expected difference in the value of $\rho$ it can be
seen that the parametric dependence of $\rho$ and $\langle k \rangle$ in both brain and Ising
networks allows us to confidently study
and compare networks of a given $\langle k \rangle$.

Next we looked at the degree variance, $\sigma^2_k=\langle k^2
\rangle-\langle k \rangle^2$, plot as a function of the average
degree in Fig.~\ref{fig2}(b) and (d). Note that the brain and the
Ising data at the critical temperature share a similar
$\sigma_k^2$-$<k>$ functional dependence, a fact that suggest
potential similarities between the degree distributions of the
brain and Ising model at $T_c$, an aspect that will be explored in
the next section. The dashed line corresponds to a Poisson
distribution ($\sigma^2_k$=$\langle k \rangle$), indicating that
none of the networks seems to obey a Poisson degree distribution.

Next we computed and compared some of the network's most basic
properties. These include the network's clustering coefficient $C$,
estimating the number of mutual connections, the average path
length $L$, defining  the average number of steps along the
shortest paths for all possible pairs of network nodes. Another
property is the diameter $D$, which is defined as the maximal
distance between any two nodes in the network. Also similar
properties $C_{ran}$, $L_{ran}$,$D_{ran}$ are computed for an
equivalent random network rewired as described in~\cite{maslov}.
Table I contains the results of these calculations for networks
defined with a positive $\rho$ and Table II the results for
negatively correlated networks. The Ising data correspond to the
already described generated networks at different temperatures ,
and the brain network corresponds to the subject whose correlation
is plotted with dashed line in Fig.~\ref{fig1} (Internal Code
Subject01). In all cases, we impose similar $\langle k \rangle$,
by choosing the appropriate $\rho$ value given by the results in
Fig. 2 a and c.

The most remarkable result in Table I is the fact that the Ising
model network become small world only at $T_c$. As expected by the
divergence of correlations at criticality, the diameter of the
network is seen here to grow from a few nodes to a length towards
the lattice maximum (i.e., $\sim \sqrt{(L/2)^2+(L/2)^2}$). At the
same time, the average minimum path $L$ only doubles. Also, it is
only at criticality that the clustering coefficient, which is
descriptive of the ''local'' connectivity, grows several orders of
magnitude compared with either sub or supercritical networks. It
is interesting to realize that the network became small world at
$T_c$, not by adding short cuts to a previously ordered lattice,
as in the Watts \& Strogatz \cite{WS} scenario. Instead, here it
seems as if the disordered small blobs coalesce (or group) at
$T_c$ producing an increase of $C$ while increasing the $D$ and
maintaining the same $L$. The other important point to remark on
here is the fact that a purely dynamical property (i.e.,
criticality) is able to dramatically change the network
properties. This has deep relevance to brain function, since these
emergent properties are directly related to the efficiency of
information transport in the network. Turning to the brain data in
Table I, it can be seen that it compares well with the Ising data
at $T_c$, and also that it is a small world network, something
reported  earlier for subjects performing minimal
tasks~\cite{eguiluz}. The data also agree extremely well with a
very recent report for resting state networks\cite{vanden}.

\begin{table}
\begin{center}
\begin{tabular}{|cccccccccc|} \hline
& & & &  & Ising network & & & &
\\ \hline
%& & &positive network  & & & negative network  & \\
T & $\rho^+$ & N & $\langle k \rangle$ & $C$ & $L$ & $D$ &
$C_{ran}$ & $L_{ran}$ & $ D_{ran}$ \\\hline
2.0  & 0.1 & 40000 & 133 & 0.065 & 2.72 & 4 & 0.0054&2.61 &4 \\%40000 \\
2.3  & 0.3 & 40000 &127 & 0.516 & 6.83 & 31 & 0.048 & 2.71 & 5  \\%39987 \\
3.0  &  0.09 & 40000 &128 & 0.064 & 2.73 & 4 & 0.0034& 2.65 &4
\\\hline
\hline%40000 \\ \hline
& & & & & Brain network & & & & \\ \hline
 & 0.623 & 26985 & 128 &  0.4536 & 4.4 & 13 & 0.061 & 2.62 & 5 \\ \hline
\end{tabular}
 \caption{Average statistical properties of the positively correlated brain and Ising networks.}
 \end{center}\label{tabla1}
\end{table}

The properties of negatively correlated networks necessarily
mirror the networks already discussed, with one important
exception. Considering the Ising model for description sake, in
the negatively correlated network edges connect spins of opposite
signs. Thus the clustering coefficient $C$ is by definition zero
since two spins mutually opposite are necessarily similar. For the
case of the brain, of course, although the details are different
the mechanics are the same.
\begin{table}
\begin{center}
\begin{tabular}{|cccccccccc|} \hline
& & & &  & Ising network & & & &
\\ \hline
T & $\rho^-$ & N & $\langle k \rangle$ & $C$ & $L$ & $D$ &
$C_{ran}$ & $L_{ran}$ & $ D_{ran}$ \\\hline
2.0 &-0.0575 & 29370 & 26 & 0.0105 &  3.24 & 5 & 0.013 & 3.38 & 8  \\%25447\\
2.3 & -0.38 &  4784 & 26 & 0& 4.17 & 13 & 0.095 & 2.95 & 7   \\% 7070 \\
3.0  & -0.105 & 40000 & 23 & 0.00003 & 3.72 & 6 & 0.0007 &3.7
&6\\\hline

& & & & & Brain network & & & & \\ \hline

 %39999\\\hline \hline
 & -0.71 & 1684 & 27 & 0& 3.54 & 11 & 0.14 & 2.69 & 5 \\\hline

\end{tabular}
\caption{Average statistical properties of the negatively
correlated brain and Ising networks.}
 \end{center}\label{tabla2}
\end{table}

\begin{figure}
\begin{center}
\includegraphics[angle=0,width=0.3\textwidth,keepaspectratio,clip]{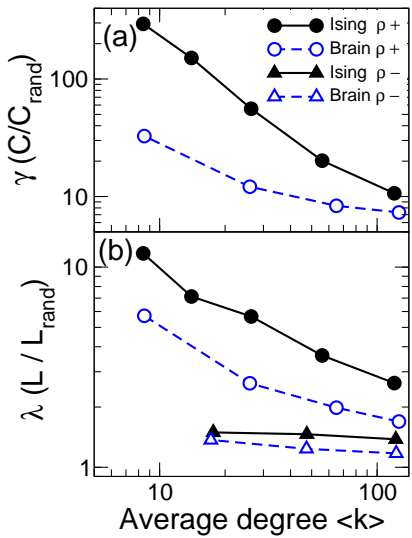}
\end{center}
\caption{Normalized network clustering (Panel a) and path length
(Panel b) as a function of average degree. Solid lines denote
results for the Ising model at $T_c$  and dashed lines the results
for the brain.}\label{fig3}
\end{figure}

Of course the above mentioned results depend on the values of
chosen $\langle k \rangle$. Nevertheless, we have verified that at
least in the range of $\langle k \rangle < 130$ the
correspondence between the Ising model at $T_c$ and the brain data
continues.  As the mean degree $\langle k \rangle$ increases the
clustering increases and the average path length diminishes, mainly
because we are adding connections to the network. A similar
dependence of these average statistical quantities on $\langle k
\rangle$ was recently reported in \cite{vanden} for positively
correlated brain fMRI networks. To summarize the dependence of
 these quantities on $\langle k \rangle$, Fig.~\ref{fig3}, shows
the normalized clustering, $\gamma=C/C_{ran}$(panel (a)), and
average path length, $\lambda=L/L_{ran}$ (panel (b)), as a
function of mean degree ($\langle k \rangle$). For the positively
correlated networks,  the relative large values of $\gamma$
together with relatively small $\lambda$ indicate that both
networks display small world properties. For the negatively
correlated network, the clustering is zero ($\gamma=0$), and
$\lambda$ is near one and notoriously almost insensitive to $\langle k
\rangle$.

The results described so far show that the dynamics of the Ising
model at $T_c$ as captured by the correlation networks exhibit
average statistical properties resembling those observed in the
brain networks at resting conditions. In the next section, the
extent of these similarities is further expanded to other network
topological features.

\subsection{Degree distribution and degree correlations}
\begin{figure}
\begin{center}
\includegraphics[angle=0,width=0.80\textwidth,keepaspectratio,clip]{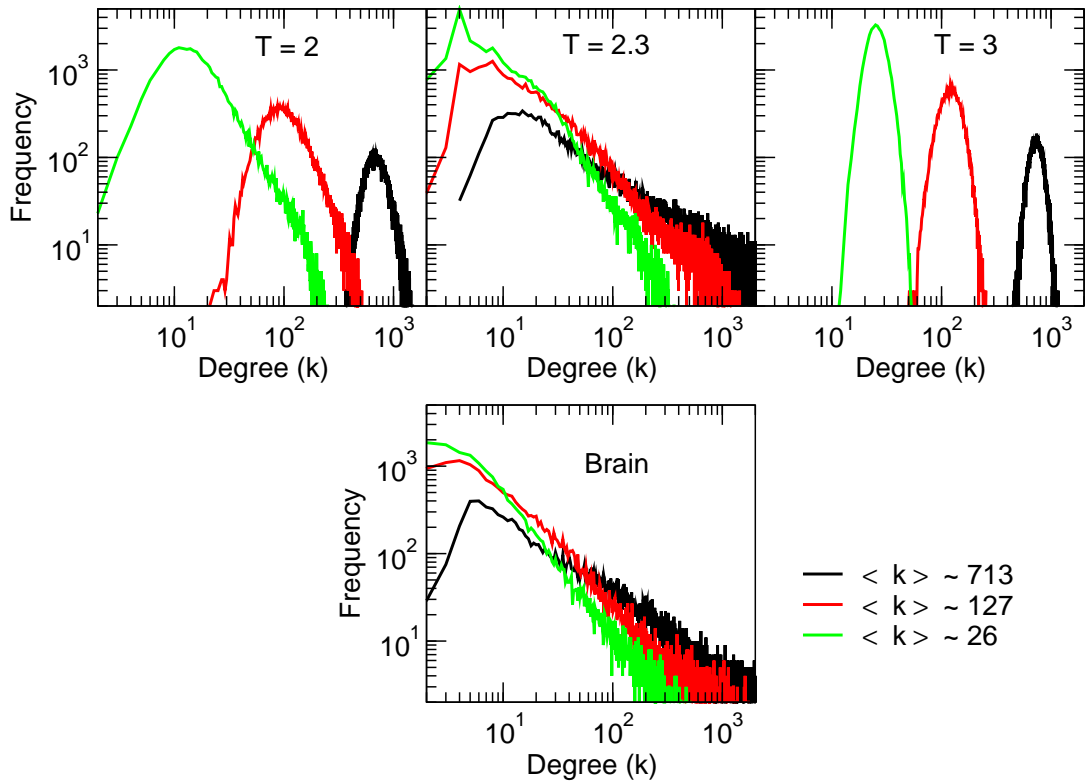}
\end{center}
\caption{Degree distribution for positively correlated networks.
Top three panels depict the degree distribution for the Ising
networks at $T=2$, $T=2.3$ and $T=3$ for three representative
values of $\langle k \rangle \approx 26$, $127$, and $ 713$.
Bottom panel: Degree distribution for positively correlated brain
network for the same three values of $\langle k \rangle$. }
\label{fig4}
\end{figure}

In this section we analyze the distribution of the networks edges
and their mutual correlation.  First we analyze and compare the
degree distribution $P(k)$, shown in Fig.~\ref{fig4}. The plots in
the top three graphs correspond to the degree distribution for the
Ising model at the three temperatures. Each of the three curves in
each graph corresponds to networks with average degree $\langle k
\rangle \approx 26, 127$ and $713$ imposed  by choosing
appropriate values of $\rho^+$ as done before.  The bottom graph
in Fig.~\ref{fig4} shows the degree distribution for the brain
network. As anticipated, the networks extracted from the Ising
model show a dramatic change at $T_c$. At criticality the degree
distribution exhibits a long tail that persists for all $ \langle k
\rangle$ explored. The power law exponent of the degree distribution and the
$ \langle k \rangle$ are connected, which is not surprising. The
bottom panel shows the same analysis for the brain network, which
exhibits all the relevant features seen for the Ising model at
$T_c$. Besides the agreement on the gross features of the
distribution, it is even possible to identify, a noticeable
maximum at $k=4$ (or $k=6$ in the brain), which is trivially
related to the number of the nearest neighbors (this is especially
notorious at relatively large $\rho^+$, see for instance Fig. 2 of
Eguiluz et al. \cite{eguiluz}).
The power law behavior in the tail of the degree distribution for
positive correlation networks was already reported in brain fMRI
of human subjects performing minimal attention task \cite{eguiluz}
and recently in an extensive study in subjects during resting
state\cite{vanden}.

\begin{figure}
\begin{center}
\includegraphics[angle=0,width=0.80\textwidth,keepaspectratio,clip]{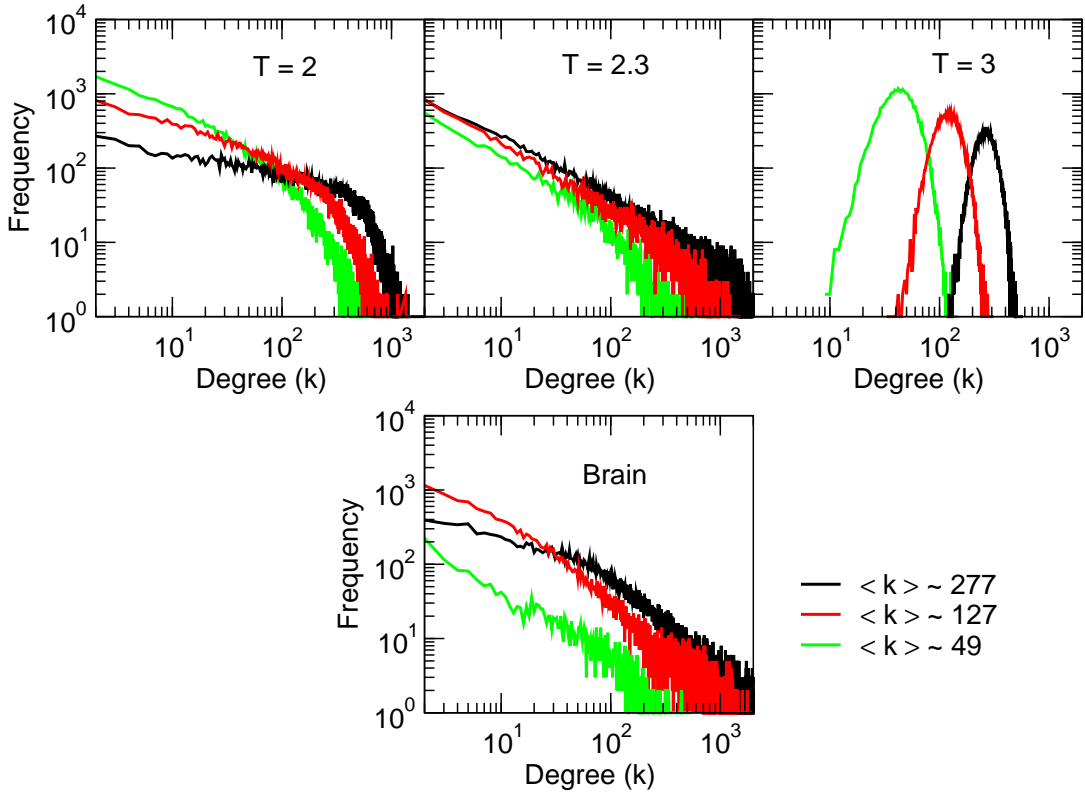}
\end{center}
\caption{Degree distribution for negatively correlated networks.
Top three panels depict the degree distribution for the Ising
networks at $T=2$,$T=2.3$,$T=3$ respectively for two
representative values of $\langle k \rangle \approx 49$ ,127 and
277. Bottom panel: Degree distribution for negatively correlated
brain network for the same three values of $\langle k \rangle$.}
 \label{fig5}
\end{figure}

Turning the attention to the negatively correlated networks, in
Fig.\ref{fig5} the degree distribution is shown for three
temperatures and two values of $\langle k \rangle \approx$ 49, 127
and 277. As seen before with other network features, there is also
a qualitative change at $T_c$ in the tail of the degree
distribution which follows a power law. The bottom panel of
Fig.\ref{fig5} shows the degree distribution for the brain
negative correlation network which presents similar features as
those seen in the Ising at $T_c$. The average for five subjects is
shown in Fig. 6.

\begin{figure}
\begin{center}
\includegraphics[angle=0,width=0.35\textwidth,keepaspectratio,clip]{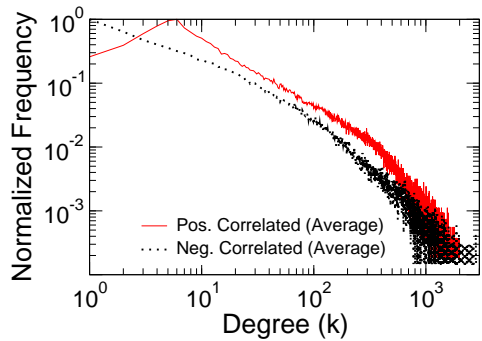}
\end{center}
\caption{Brain network's average degree distribution computed from
five volunteers for $\langle k \rangle \approx 127$.}
 \label{fig6}
\end{figure}

\begin{figure}
\begin{center}
\includegraphics[angle=0,width=0.60\textwidth,keepaspectratio,clip]{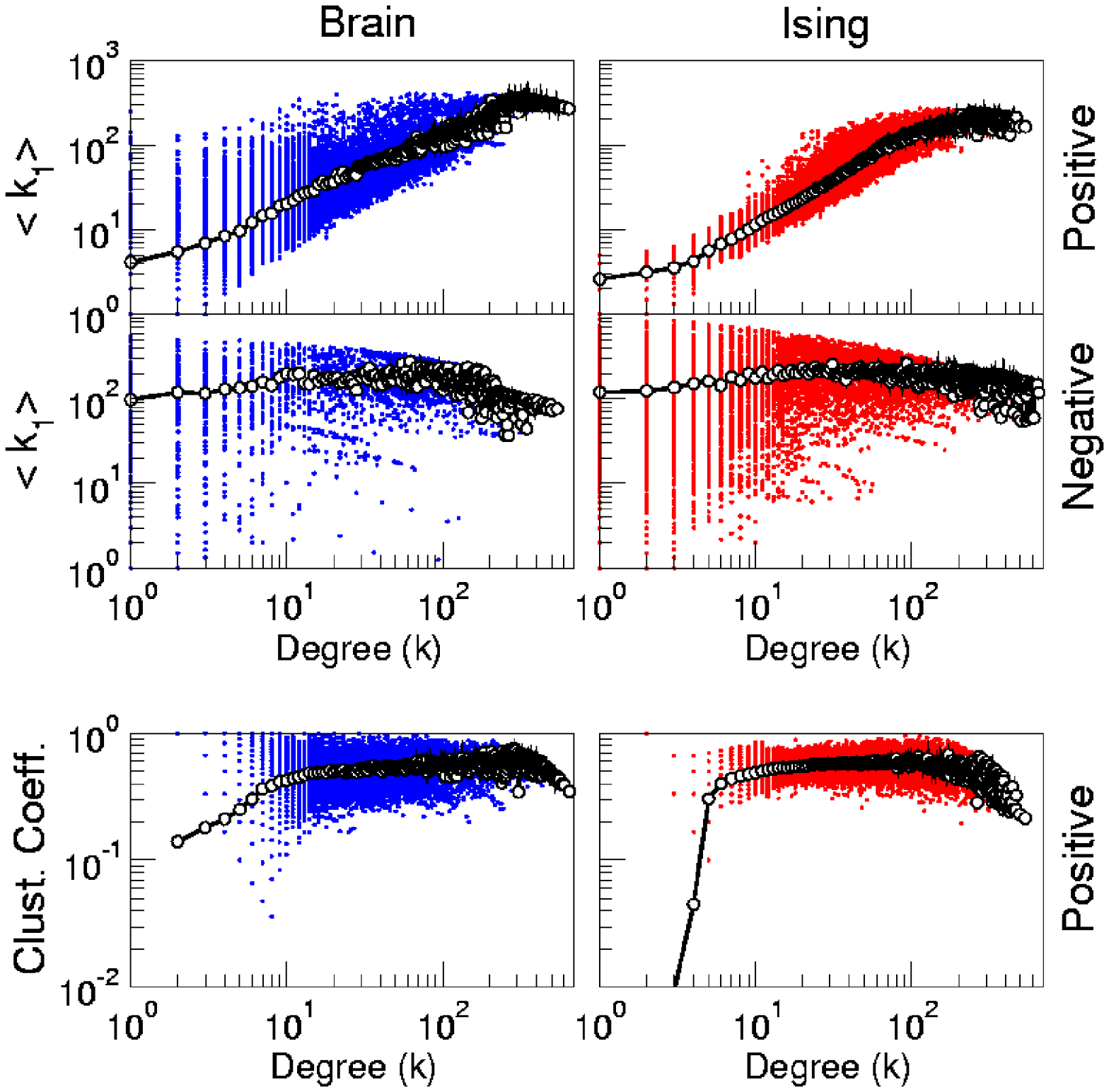}
\end{center}
\caption{Top four panels: Neighbor Degrees Correlation. Plot of
the nearest neighbor degree, $\langle k_1\rangle $, as a function
of own degree $k$ for the two type of networks extracted from the
brain (left) and from the Ising model at $T_c$ (right).  Bottom
panels: Clustering, $C$, as a function of the degree $k$ for
positively correlated networks extracted from the brain (left) and
from the Ising model at $T_c$: In all plots dots represent
individual nodes, and empty circles joined by lines represent
averages. Positively correlated networks correspond to $\langle
k\rangle \approx 26$ and negatively correlated network correspond
to $\langle k\rangle \approx 49$.} \label{fig7}
\end{figure}

In some complex networks, a node's degree and its neighbors'
degree can be related. The correlation between node degree and
neighbor degrees, as well as the dependence of other measures on a
node's degree are investigated in Fig. 7. The bottom panels of
Fig.~\ref{fig7} illustrate the relation between the clustering,
$C(k)$, and the degree for positively correlated networks. In can
be seen that for the most part $C(k)$ is independent of $k$ in
both the brain and the Ising model networks at $T_c$.

The top four panels of Fig.~\ref{fig7} compare the nearest
neighbor degree, $\langle k_1\rangle $, as a function of own
degree $k$ for the two types of networks extracted from the brain
and from the Ising model at $T_c$. In both positively correlated
networks one can see the so called assortative property, by which
highly connected nodes tend to be connected with highly connected
neighbors. The presence of this feature, first described in
Eguiluz et al\cite{eguiluz}, can now be understood considering
that is linked with the dense domains of equally oriented spins
(or voxels). Sites located deep into the bulk of the domain then
will have a larger degree, and by the same reasoning sites located
in the domain's periphery will result in nodes with smaller
degree. Then the assortative property is, in this context,
trivially related to the geographical location of each node.
Having clarified this, it follows that in the case of the
negatively correlated networks, a node's neighbors won't be
affected in the same way by their location, resulting in the
degree independence seen in Fig. 7.

The same reasoning can reconcile apparently conflicting results
(reviewed in\cite{bassetreview}), in which brain network degree
distributions were found not to be scale free. In this work the
fMRI time series inside relatively large predefined cortical areas
were first averaged. Then the correlation between these averages
(a few dozen for the entire brain) were used to define the
networks. From the discussion above it is clear that the averages
remove the main source of the long tails we observe here. The
local averaging precludes the possibility of observing these details.
A gross comparison would be the effect of recalculating the degree
distribution of the United States' airline traffic, which is known
to be scale free, by no longer considering airports as its
nodes, but rather averaging traffic between entire states. Of course,
this averaging obscures the hubs and prevents the observation of
such scales.

In passing it should be mentioned that a discussion of the relevance
of this spatial aspect over spurious clustering coefficients was
reported recently \cite{tsonis} for networks constructed from
pressure levels representative of the general circulation (wind
flow) of the atmosphere.

\subsection{Back to plain correlations}

Previous sections demonstrate striking similarities between the
correlation properties of the brain and the Ising model at $T_c$.
This was done comparing the correlation networks, a technique that
as commented in the introduction allows for a compact description.
However, to be consistent those similarities should be evident by
looking at plain correlations as discussed next.

\begin{figure}
\begin{center}
\includegraphics[angle=0,width=0.3\textwidth,keepaspectratio,clip]{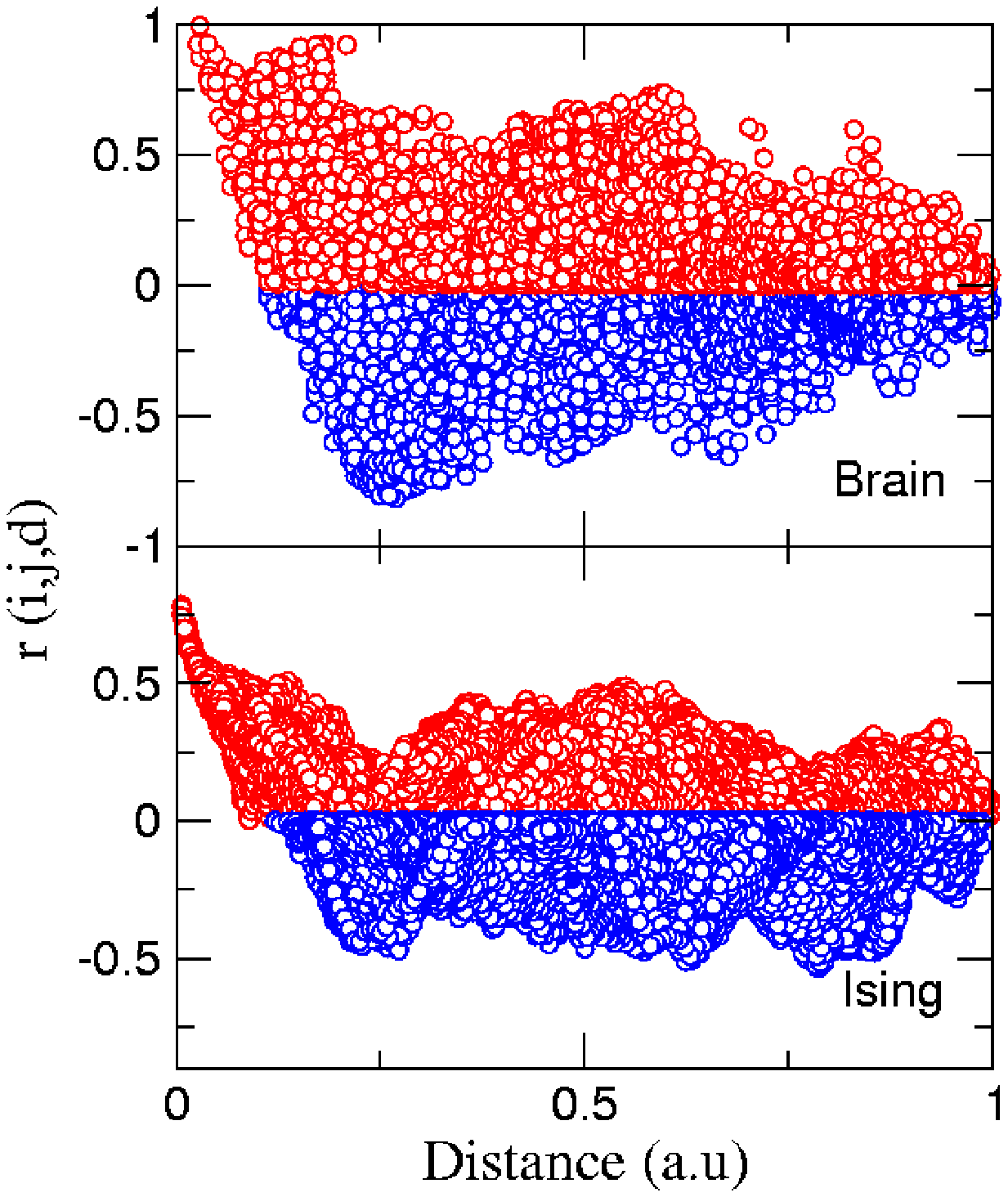}
\end{center}
\caption{Typical brain (top) and Ising (bottom) correlation
profiles. Correlations (i.e., Eq. 2.) are computed between the
site with the largest degree and the rest of the time series and
plotted at its respective normalized distance} \label{fig8}
\end{figure}

In previous work we reported already that on the \emph{average}
correlations in the brain decay very slowly with
distance\cite{eguiluz}. We revisit this aspect here both for the
brain and the Ising model. This is shown in Fig.~\ref{fig8}.
%%%%%%%%%%%%%%%%%%%%%%%%%%%%%%%%%%%%%%%%%%%%%%%%
%%%%%%%%%%%%%%%%%%%%%%%%%%%%%%%%%%%%%%%%%%%%%%%%%%%%
The first observation is that significant correlations extend to the
length of the system. Near the origin, there is a notorious bias
toward positive correlations, followed by a somewhat rough
landscape of both positive and negative correlations. In the case
of the brain, the peaks of this landscape reveal areas with
common anatomical and (probably) functional properties.

It is important to recognize that the valleys of negative
correlation may or may not be related to negative interactions.
Since there are no negative interactions in Eq. 1, it is clear
that in the Ising model the emergence of significant negative
correlations is a collective effect arising only at the critical
point. In the brain the situation is not that clear. Nevertheless
there is a pervasive preference to link anti-correlated dynamics
with negative interactions. For instance it is commonly heard that
``In the human brain, neural activation patterns are shaped by the
underlying structural connections that form a dense network of
fiber pathways linking all regions of the cerebral
cortex''\cite{SpornsPlos}. This mind set, which equates dynamics
with structure, is so entrenched in brain science that it always
seems reasonable to search for the connections responsible for any
given dynamical pattern. The results discussed here suggest that
as a change in the temperature can lead to the emergence of
correlations in a substrate that lacks such ability, the brain
cortex can be operating in the same way. In fact the puzzle of how
the brain can be highly coherent over long lengths have prompted
some authors to postulate even non-classical explanations, while
the results here seems to suggest that if the brain is at
criticality such coherence can be achieved naturally.

The second observation is that the fraction of sites which have
positive correlations is about the same as the fraction of sites
with negative correlations. Again, this feature can be easily
understood for the Ising model at criticality, but it is hard to
reconcile for the brain, unless a critical scenario is invoked.
This finding might have deep implications. For instance, we have
recently reported\cite{baliki2} that although such a balance is
maintained in healthy individuals, it is disrupted in some
pathologies. Specifically, the disruption found is a reduction of
the number of anti-correlated sites, compared with normal
conditions, somewhat analogous to subcritical temperatures in the
Ising model, a situation dominated by equally oriented domains.

\subsection{Brain Functional Connectivity vs Collectivity}
Probably is worth to place the present results in the context of
current brain imaging approaches. The literature specialized on
the analysis of brain neuro-imaging time-series includes a very
productive chapter of functional connectivity, dedicated to
formalize findings on a cohesive picture. Three basic concepts in
this area are: brain functional connectivity, effective
connectivity and structural connectivity
\cite{friston,spornsconnectome,horwitz}. The first one ``is
defined as the correlations between spatially remote
neurophysiological events''\cite{friston}. Per se, the definition
is a statistical one, and  ``is simply a statement about the
observed correlations; it does not comment on how these
correlations are mediated''\cite{friston}. The second concept,
effective connectivity is closer to the notion of a neuronal
connection and ``is defined as the influence one neuronal system
exerts over another''. Finally, the concept of structural or
anatomical connectivity refers to the identifiable physical or
structural (synaptic) connections linking  neuronal elements.

These three concepts, intentionally or not, emphasize the
connection between brain elements. And it is despite of cautionary
comments  emphasizing explicitly that ``depending on sensory
input, global brain state, or learning, the same structural
network can support a wide range of dynamic and cognitive
states''\cite{spornsconnectome}. In this regard, the present
results are specific examples of the emergence of nontrivial
collective states over an otherwise trivial regular lattice (i.e.
the Ising's structural connectivity). It is clear that if the
Ising's collective states have a counterpart in the brain they can
not be adequately described in the framework of connectivity,
rather it would be more appropriate to define another framework in
terms of brain functional \emph{collectivity}.  A pedestrian
starting point, would be to review instances in which the data
from the brain structural connectivity and the functional
correlations disagree, as indications of collective phenomena.

\section{SUMMARY AND DISCUSSION}

In this work statistical properties of brain correlation networks
have been compared with those of networks derived from the 2D
Ising model. The main finding here is that at the proper
temperature Ising networks and brain networks are
undistinguishable from each other. Their main topological
properties and even more refined features of network structure
including degree distribution, neighbor degree and clustering
correlations with own degree, all behave in the same manner.

The biologically most relevant lesson is related to the well
known central result in critical phenomena. Namely that the
dynamics of a system near a critical point include spatiotemporal
patterns correlated and anti-correlated over long distances,
despite having only nearest neighbor positive interactions. The
similarities exposed by the comparison made in this paper suggest
that collective dynamics  with similar mechanics are present in the
brain.

Nevertheless, the main point is not that a simple model completely
orphan of neural details is able to replicate the experimental
observations. The main point is that the model gets the correct
phenomenology without explicitly plugging components for such
phenomenology into its equations. Whatever ends up replicating the
observations, it is a collective effect that only happens at a
certain temperature. Naturally, this runs against common sense in
brain science because, as discussed in previous sections, the
prevailing mind set implies that if two brain regions act in some
coherent way, there must be a direct connection between them. The
commonly held view seems to see the brain as a low temperature
system, while we all have first hand experience with
brain/behavior changes that are known to be the action of
relatively broad and completely unspecific factors. We suggest
that some type of global changes (e.g., mood, arousal, attention,
etc) might be brought about in the same way that coherent domains
arise at the critical temperature. Despite the relative abundance
of approaches, including sophisticated ones, as far as we know, no
neural model relies solely on this kind of dynamical transition as
a mechanism to produce different behaviors.

In summary we have compared networks derived from the fMRI signal
of human brains with similar networks extracted form the Ising
model. We found that near the critical temperature the two
networks are indistinguishable from each other for most relevant
statistical properties. These results are interpreted here as
lending support to the conjecture that the dynamics of the
functioning brain is near a critical point.

\section*{Acknowledgments}
This work was supported by NIH NINDS NS58661.

%%%%%%%%%%%%%%%%%%%%%%%%%%%%%%%%%%%%%%%%%%%%%%%%%%%%%%


\begin{thebibliography}{9000}
\bibitem{werner}G. Werner. \emph{Journal of Physiology - Paris}, \textbf{101}, 273--279
(2007).
\bibitem{fox_07} M.D. Fox and M.E. Raichle. \emph{Nat. Rev. Neurosci.}, \textbf{8}, 700 (2007).
\bibitem{raichle_01} M.E. Raichle, A.M.  MacLeod, A.Z. Snyder, W.J. Powers,
D.A. Gusnard, G.L. Shulman. \emph{Proc. Natl. Acad. Sci. U.S.A.},
\textbf{98}, 676 (2001).
\bibitem{raichle_06a} M.E. Raichle ME. \emph{Science}, \textbf{314}, 1249 (2006).
\bibitem{greicius_03} M.D. Greicius, B. Krasnow, A.L. Reiss,
V. Menon. \emph{Proc. Natl. Acad. Sci. U.S.A.}, \textbf{100}, 253
(2003).
\bibitem{Stam}C.J. Stam . \emph{Neurosci Lett}, \textbf{355} 25–8 (2004).
\bibitem{vanden} M. P. van den Heuvel, C.J. Stam. M. Boersma and H. E.
Hulsshoff Pol, \emph{NeuroImage} (2008), in press.

\bibitem{Salvador} R. Salvador, J. Suckling,
M.R. Coleman, J.D Pickard, D. Menon, E.T. Bullmore, \emph{Cerebral
Cortex} \textbf{15}, 1332-1342, (2005).

\bibitem{Bullmore} S. Achard and  E.T Bullmore \emph{PLoS Computational Biology }\textbf{3}, 0174 (2007).
\bibitem{eguiluz} V.M. Eguiluz, D.R. Chialvo, G. Cecchi, M. Baliki, A.V. Apkarian
\emph{Phys. Rev. Lett.} \textbf{94}, 018102 (2005).

\bibitem{SpornsPlos}P.  Hagmann, L.  Cammoun, X. Gigandet,R.  Meuli, C.J. Honey, et al.
 \emph{PLoS Biology} \textbf{6} e159 doi:10.1371/journal.pbio.0060159(2008)

\bibitem{Park2008}C-H Park, S.Y. Kima, Y-H. Kimb, K. Kimc, \emph{Physica A} \textbf{387} 5958–5962
(2008).
\bibitem{Achard} S. Achard, R. Salvador, B. Whitcher, J. Suckling, and E. T.
Bullmore, \emph{J. Neurosci.} \textbf{26}, 63 (2008).

\bibitem{fox_05}M.D. Fox, A.Z. Snyder, J.L. Vincent, M. Corbetta, D.C. van Essen, M.E. Raichle.
\emph{ Proc. Natl. Acad. Sci. U.S.A.} \textbf{102}, 9673 (2005).


\bibitem{baliki2} M.N. Baliki, P.Y. Geha, A.V. Apkarian, D.R. Chialvo. \emph{J. Neuroscience.}, \textbf{28}(6), 1398 (2008).
\bibitem{chialvo2004} D.R. Chialvo. \emph{Physica A} \textbf{340},756 (2004).
\bibitem{chialvo2007} D.R. Chialvo.  \emph{AIP Conference Proceedings}, \textbf{887}, 1 (2007).
\bibitem{chialvo2008} D.R. Chialvo, P. Balenzuela, D. Fraiman. \emph{AIP Conference Proceedings}, \textbf{1028} 28
(2008).

\bibitem{vincent_07} J.L. Vincent, G.H. Patel, M.D. Fox, A.Z. Snyder, D.C. Van Essen, M. Corbetta, M.E. Raichle \emph{Nature}, \textbf{447}, 83 (2007).

\bibitem{Stamreview1}C.J. Stam  and J.C. Reijneveld \emph{Nonlin Biomed Phys} \textbf{1} 3
(2007).

\bibitem{Stamreview2} J.C. Reijneveld, S.C. Ponten , H. W. Berendse , C. J. Stam.
\emph{Clinical Neurophysiology}, \textbf{118} 2317–2331 (2007)


\bibitem{Ising}E. Ising, \emph{Z. Phys.}, \textbf{31}, 253–258
(1925).

\bibitem{jezzard}P. Jezzard, P. Mathews, S.M. Smith \emph{Functional
MRI: An introduction to methods}, (Oxford University Press, 2001).

\bibitem{Metro} N. Metropolis, A.W. Rosenbluth, M.N. Rosenbluth, A.H. Teller, and E. Teller.
\emph{Journal of Chemical Physics}, \textbf{21} 1087 (1953).

\bibitem{Tobo} H. Gould and J. Tobochnik. \emph{An introduction to computer simulations methods} (Addison Wesley
1996).

\bibitem{Basset} D.S. Bassett, A. Meyer-Lindenberg, S. Achard, T. Duke, E.
T. Bullmore, \emph{Proc. Natl. Acad. Sci. U.S.A.} \textbf{103},
19518 (2006).
\bibitem{maslov} S. Maslov, K. Sneppen, U. Alom,  \emph{Handbook of graphs and networks},
S. Bornholdt and H.G Schuster (Eds.) (Wiley-VCH and Co., Weinheim,
2003)

\bibitem{problem}This is a fundamental brain puzzle, a stability problem which still remains
unsolved. Solutions that will work in relatively small, linear and
delayless systems, eventually breakdown when realistic delays are
added, sizes are scale up and nonlinear aspects considered. A
quantitative discussion of these issues can be found in page 104
of \cite{abeles}.

\bibitem{abeles} M. Abeles  \emph{Corticonics. Neural circuits of the cerebral
cortex}, (Cambridge University Press, 1991).

\bibitem{WS} D.J. Watts and S.H. Strogatz \emph{Nature}, \textbf{393} 440
(1998).

\bibitem{bassetreview}D.S. Bassett  and E.T. Bullmore  \emph{Neuroscientist} \textbf{12} 512–523
(2006).

\bibitem{tsonis} A.A. Tsonis, K. L. Swanson, G. Wang \emph{Physica A} \textbf{387}
5287-5294 (2008).

\bibitem{friston} K.J. Friston
%Functional and effective connectivity in neuroimaging: a synthesis.
\emph{Hum. Brain Mapp.} \textbf{2} 56–78 (1994).

\bibitem{horwitz}B. Horwitz
%The elusive concept of brain connectivity.
\emph{Neuroimage}\textbf{19} 466-470 (2003).

\bibitem{spornsconnectome} O. Sporns, G. Tononi, R. Kotter
%The human connectome: A structural description of the human brain.
\emph{PLoS Comput. Biol.} \textbf{1} 245-251 (2005).

\bibitem{honey}Honey CJ, Kotter R, Breakspear M, Sporns O. \emph{Proc Natl Acad Sci U S A}, \textbf{104}
10240–10245 (2007).



\end{thebibliography}
\end{document}